# Exploratory Learning Environments for Responsible Management Education Using Lego Serious Play

Vasilis Gkogkidis [a,*] & Nicholas Dacre [a]

[a] University of Southampton Business School, Southampton, UK
[*] Corresponding Author: v.gkogkidis@southampton.ac.uk

## Abstract

Research into responsible management education has largely focused on the merits, attributes, and transformation opportunities to enhance responsible business school education aims. As such, a prominent part of the literature has occupied itself with examining if responsible management modules are inherently considered a non-crucial element of the curriculum and determining the extent to which business schools have introduced such learning content into their curriculum. However, there has been scant research into how to apply novel teaching approaches to engage students and promote responsible management education endeavours. As such, this paper seeks to address this gap through the development of a teaching framework to support educators in designing effective learning environments focused on responsible management education. We will draw on constructivist learning theories and Lego Serious Play (LSP) as a learning enhancement approach to develop a pedagogical framework. LSP is selected due to its increasing application in learning environments to help promote critical discourse, and engage with highly complex problems, whether these are social, economic, environmental, or organisational.

# Introduction

Business schools play a vital role in educating and shaping the mindsets of future global leaders, as they aim to offer a rich environment of deep analysis, entrenched engagement, and thought-provoking discourse (Harrison et al., 2007). Their role also extends beyond the confines of the academic world as they have increasingly become embedded in their respective national economies (Thorpe & Rawlinson, 2013) where they help foster innovations by collaborating with industry, academic experts, and thought-leaders (Minshall & Wicksteed, 2005). Furthermore, as the relationship between society, the environment, and business becomes ever-more intertwined, business schools and educators have an increasingly intricate role in signifying the complexity of these elements to help develop socially responsible professionals that can draw on an ethical management ethos, and apply it to future practice (Dyllick, 2015; Godemann et al., 2014).

The concept of responsible management education promotes such aims, with key principles underpinned by organisations such as the United Nations (Stefanova & Stefanova, 2013). However responsible management education has hitherto largely focused on what constitutes elements of corporate social responsibility (CSR), and how these should be embedded into the curriculum (Aragon-Correa et al., 2017; Dyllick, 2015). In part this has been driven by a delineated interpretation of responsible management education topics.

For example, there has been a focus on content which is salient to execution, and therefore driven by the analysis of the plethora of social, economic, and environmentally aware subjects which may bear relevance to business school education contexts (Muff, 2013). These include elements such as sustainable supply chains, CSR and the circular economy (Snelson-Powell et al., 2016). Nonetheless, business schools have been criticised for their lack of progress in recognising the importance of these responsibility-aware topics (Dyllick, 2015).

In turn this has focused institutions and educators on the premise of content-driven responsible management education, with limited recognition of the processes of engagement and delivery. Notwithstanding the importance of this focus, business schools and educators are facing complex teaching and learning challenges in order to deliver these subjects in a way that matters most for their audience, in this case students, who represent future practitioners (Dacre et al., 2019; Ojiako et al., 2011).

As such, by bringing together constructivist learning theories like exploratory learning environments (Duckworth, 2006; Rick & Lamberty, 2005), organisational sustainability teaching frameworks (Stubbs & Cocklin, 2008) and empirical research on the use of the Lego Serious Play (LSP) methods in educational settings (James, 2013; Kurkovsky, 2015; Mccusker, 2014), this research will aim to conceptualise LSP as an innovative teaching method that utilises Lego bricks to improve student engagement and participation, create exploratory teaching environments that can support learning, and help shape responsible organisational leaders.

## Literature Review

Responsible management education can be defined as the educational practices aiming to facilitate students' learning on issues like CSR, the social impact of organisations on the larger societies they are embedded in and the impact organisations have on the preservation of the environment (Forray & Leigh, 2012). The larger issues of how to manage organisations ethically and how these impact everyday decisions and practice are also included in the responsible management curriculum (Godemann et al., 2014). Discourse calling for business schools to promote a more responsible management curriculum has increasingly gathered support in management education literature, but adapting the curriculums of business schools to this has proved to be a more challenging task than initially expected (Parkes et al., 2017; Rasche et al., 2013; Rasche & Gilbert, 2015).

One of the main issues around the implementation of responsible management curriculum is the fact that business schools have not made such courses mandatory for all students, with 75% of these modules offered as electives, thus remaining detached from core disciplines (Rasche et al., 2013). Similar conclusions were drawn by the Principles for Responsible Management Education (PRME, 2014) indicating that, even though there have been the curriculum additions of CSR and sustainability modules, they are not yet the central focus of business management education, thus failing to effectively address the challenge in encouraging more responsible future organisational leaders (Rasche & Gilbert, 2015).

Recognising the need for improved student engagement and participation, teaching staff in higher education have in recent years introduced a plethora of innovative playful pedagogical tools in the classroom, such as board games, digital games, simulations and role-playing activities (Feinstein et al., 2002; Wyss-Flamm & Zandee, 2001). Such endeavours seek to increase student engagement and facilitate active learning by involving students in the educational process more, compared to instructional teaching approaches (Dacre et al., 2015, 2018; Gkogkidis & Dacre, 2020).

LSP is also increasingly being recognised as an innovative and adaptable method for teaching and learning practices in higher education (Peabody & Noyes, 2017). A growing body of literature examines the merits and application of LSP as a teaching method with studies reporting improvement in student engagement, participation, knowledge co-creation and knowledge retention (Grienitz & Schmidt, 2012; James, 2013; Mccusker, 2014).With LSP having successfully been utilised in different knowledge domains such as computer science (Kurkovsky, 2015) and management education (Grienitz & Schmidt, 2012) this paper suggests that LSP can enable educators to embed the values of constructivist learning theories into their teaching practices and operationalise exploratory learning environments to enhance student engagement and participation.

Rick & Lamberty (2005) define exploratory learning environments as educational arrangements and activities that facilitate the learners' ability to construct knowledge connected to the subject matter through student led reflective exploration. Exploratory learning environments enable student participation in learning processes by offering their own ideas and interpretation of the knowledge under discussion. Educators adopt a facilitative rather than prescriptive role aiming to create educational experiences based on values of constructivist learning theories (Bruner, 1961).

Similar perspectives have been shared by later scholars such as King (1993), arguing for a shift in the role of the educator from 'sage-on-the-stage' to 'guide-on-the-side', facilitating learning rather than imposing it. LSP as a teaching method embodies constructivist learning theories emphasising exploration where "to understand is to discover, or reconstruct by discovery" (Piaget, 1972, p. 20) while at the same time there must be a recognition of knowledge being created in specific cultural contexts among educators and learners (Vygotsky, 1980). Exploratory learning environments facilitate knowledge communities where their members participate in their practices purposefully, with knowledge residing in the specific context it is being used in (Hickey & Zuiker, 2005).

Prior student knowledge and experiences, as influenced by the students' social and cultural environment, are viewed as a crucial element of the learning processes that help construct knowledge, and thus should ideally be discussed and negotiated among students and educators (Salomon, 1997). Students and educators, in engaging with responsible management curricula, aim at producing and negotiating knowledge that will inform future ethical management practices. Both undergraduate and postgraduate student cohorts in business schools come from a variety of different national and cultural backgrounds, with many of them, especially those undertaking postgraduate courses, having prior experience of management in organisations (Arbaugh et al., 2010; Jabbar & Hardaker, 2013; Tompson & Tompson, 1996).

Bringing these student experiences and knowledge at the forefront by discussing and negotiating rather than ignoring them is a challenge that LSP can assist educators with, as a methodology designed around the values of active and egalitarian participation. Finally, constructivist theories suggest that new knowledge acquired by students is added to existing knowledge schemes, mental models that keep expanding when new understandings are achieved (Hoidn, 2017). This idea is especially pertinent to responsible management education, where students with existing knowledge of management theories and practices are offered the opportunity to enhance their knowledge with responsible management frameworks, assisting the transition towards a more sustainable type of management.

# Exploratory Learning

Providing new teaching methodologies to support a better understanding of sustainability concepts in management education (Cervantes, 2007), Stubbs & Cocklin (2008) suggest a pedagogical approach where educators present students with a typology of business sustainability practices, followed by two case studies of organisations that engage in business sustainability practices.

The suggested typology describes organisations following an eco-centric approach striving to create closed-loop systems where unused materials belonging to one organisation in the system are used as an input for another organisation (Stubbs & Cocklin, 2008) and destruction of environmental resources is minimised by sharing infrastructure between organisations (Ayres & Ayres, 2002). Organisations designed around the idea of ecological modernisation aim at profitability while contributing to the wellbeing of the organisation's stakeholders, and keeping environmental impacts like pollution to a minimum (Gladwin et al., 1995).

Neoclassical organisations measure all of their activities based on their economic outcomes, where sustainability is not part of their core strategy unless they strive to increase profits by strengthening their competitive advantage, comply with legislation, address concerns and pressure from the public, or are pressured by stakeholders (Banerjee, 2001; Bansal & Roth, 2000; Shrivastava, 1995).

In their paper, Stubbs & Cocklin (2008) outline two case studies, one from the banking industry and one from the automotive industry, but do not offer a reference for these case studies. We suggest publications by Hamschmidt (2007) and Vives Gabriel (2017) which offer case studies which educators can use to teach sustainability and ethical approaches to management.

An LSP methodology offers a framework to support the delivery of learning sessions, during which participants are led through a series of exercises where they build Lego models in response to the educator´s questions (Roos et al., 2004). The aim of the Lego models is to help participants share their insights about the question at hand by using metaphors rather than to accurately represent entities found in the physical world (Grienitz & Schmidt, 2012). A single yellow Lego brick can for example represent the sun or a banana or a sandy beach. Meaning is embodied in the bricks by the creator of the model to support them in answering the question that is under discussion.

LSP realises constructivist learning theories and values by offering a framework for educators to design learning environments where participants get to participate in social processes that help them learn from their teachers and peers. Central to the LSP methodology is the idea of constructing knowledge from previous knowledge and experiences by getting workshop participants together and encouraging them to use Lego materials to make and express meaning that they share with their peers (Roos et al., 2004). LSP offers an etiquette for participants and facilitators to follow and a core process which includes four stages of facilitation: the educator poses a question, students construct a Lego model answering the question, students share the meaning of their Lego models and finally the educator and students share reflections on the meanings shared (Kristiansen & Rasmussen, 2014).

## Further Research

This research will seek to develop a framework for responsible management education, teaching and learning delivery. Participatory teaching approaches like LSP present a salient opportunity for educators delivering responsible management content to enhance student engagement and strive towards a more open and inclusive teaching paradigm (Bovill et al., 2011; Cook-Sather et al., 2014; Gkogkidis & Dacre, 2020), one that will help shape more responsible organisational leaders in the future. Therefore, in developing this research further, firstly the conceptual teaching framework using LSP will have to be further developed from both research and practice-based experience by the authors. Secondly, scalability of delivery will have to considered as an element which requires further input.

For example, the use of LSP will likely be heavily reliant on the use of materials, which are primarily Lego bricks, and require the educator's attention when teaching a manageable cohort of individuals. Thirdly, this research will outline an innovative teaching methodology that can assist educators create effective learning environments for management students. In summary, in this study will develope a framework for educators who are interested in enhancing the delivery of their responsible management education to better prepare their students for pressing future global challenges.

If management as a practice is to contribute towards overcoming a range of increasingly diverse challenges, such as the environmental crisis, the structure of the economic and financial system and broader social and business concerns (Godemann et al., 2014; PRME, 2008; PRME, 2014), it is important for tomorrow's leaders to appreciate and acknowledge the complexity of these interrelated issues.

Gkogkidis & Dacre    2020Bruner J: The Act of Discovery. *Harv Educ Rev.* 1961; 31: 21–32.

Cagle JAB, Baucus MS: Case studies of ethics scandals: Effects on ethical perceptions of finance students. *J Bus Ethics.* 2006; 64(3): 213–229. https://doi.org/10.1007/s10551-005-8503-5

Cervantes G: A methodology for teaching industrial ecology. *International Journal of Sustainability in Higher Education.* 2007; 8(2): 131–141. https://doi.org/10.1108/14676370710726607

Cook-Sather A, Bovill C, Felten P: Engaging students as partners in learning and teaching: A guide for faculty. John Wiley & Sons. 2014.

Dacre, N., Constantinides, P., & Nandhakumar, J. (2015). How to Motivate and Engage Generation Clash of Clans at Work? Emergent Properties of Business Gamification Elements in the Digital Economy. International Gamification for Business Conference (IGBC15), Birmingham, UK. https://doi.org/10.5281/zenodo.4633379

Dacre, N., Gkogkidis, V., & Jenkins, P. (2018). *Co-Creation of Innovative Gamification Based Learning: A Case of Synchronous Partnership*. Society for Research into Higher Education (SRHE), Newport, Wales UK. https://doi.org/10.5281/zenodo.4633514

Dacre, N., Senyo, P., & Reynolds, D. (2019). *Is an Engineering Project Management Degree Worth it? Developing Agile Digital Skills for Future Practice*. Engineering Education Research Network (EERN), Coventry, UK. https://doi.org/10.5281/zenodo.4637873

Duckworth E: The Having of Wonderful Ideas: And Other Essays on Teaching and Learning. Teachers College Press. 2006.

Dyllick T: Responsible management education for a sustainable world The challenges for business schools. *Journal of Management Development.* 2015; 34(1): 16–33. https://doi.org/10.1108/JMD-02-2013-0022

Eckhaus E, Klein G, Kantor J: Experiential Learning in Management Education. *Business, Management and Education.* 2017; 15(1): 42–56. https://doi.org/10.3846/bme.2017.345

Feldman HD, Thompson RC: Teaching Business Ethics: A Challenge for Business Educators in the 1990s. *Journal of Marketing Education.* 1990; 12(2): 10–22. https://doi.org/10.1177/027347539001200203
Exploratory Learning Environments for Responsible Management Education Using Lego Serious Play    9

Gkogkidis & Dacre 2020Hoidn S: Student-Centered Learning Environments in Higher Education Classrooms. In: *Student-Centered Learning Environments in Higher Education Classrooms.* Palgrave Macmillan US. 2017.

Jabbar A, Hardaker G: The role of culturally responsive teaching for supporting ethnic diversity in British University Business Schools. *Teaching in Higher Education.* 2013; 18(3): 272–284. https://doi.org/10.1080/13562517.2012.725221

James A: Lego Serious Play: a three-dimensional approach to learning development. *Journal of Learning Development in Higher Education.* 2013; 6: 1–18.

King A: From Sage on the Stage to Guide on the Side. *College Teaching.* 1993; 41(1): 30–35. https://doi.org/10.1080/87567555.1993.9926781

Kristiansen P, Rasmussen R: Building a better business using the Lego serious play method. John Wiley & Sons. 2014.

Kurkovsky S: Teaching software engineering with LEGO serious play. *Annual Conference on Innovation and Technology in Computer Science Education, ITiCSE.* 2015; 213–218. https://doi.org/10.1145/2729094.2742604

Mccusker S: Lego® Serious Play TM : Thinking about Teaching and Learning. *International Journal of Knowledge, Innovation and Entrepreneurship.* 2014; 2(1): 27–37.

Mello JA: Enhancing the international business curriculum through partnership with the united states department of commerce: The "E" award internship program. *Journal of Management Education.* 2006; 30(5): 690–699. https://doi.org/10.1177/1052562906289049

Minshall THW, Wicksteed W: University spin-out companies: Starting to fill the evidence gap.In: *A report for The Gatsby Charitable Foundation*. 2005.

Muff K: Developing globally responsible leaders in business schools: A vision and transformational practice for the journey ahead. In: *Journal of Management Development.* Emerald Group Publishing. 2013; 32(5): 487–507. https://doi.org/10.1108/02621711311328273

Ojiako U, Ashleigh M, Chipulu M, *et al.*: Learning and teaching challenges in project management. *International Journal of Project Management.* 2011; 29(3): 268–278. https://doi.org/10.1016/j.ijproman.2010.03.008
Exploratory Learning Environments for Responsible Management Education Using Lego Serious Play 11